\documentclass[a4paper,10pt,twocolumn,twoside]{article}
\pdfoutput=1
\usepackage{graphicx}
\usepackage{amsmath}
\usepackage{amssymb}
\usepackage{color} 

\usepackage{chicago}
\usepackage{url}
\usepackage{cite}
\usepackage{hyperref}
\hypersetup{
pdfauthor={Havemann; Heinz; Gl\"aser; Struck},
pdftitle={Havemann et al. (2013): Estimating Thematic Similarity of Scholarly Papers with Their Resistance Distance in an Electric Network Model},
colorlinks=true,
linkcolor=blue,
anchorcolor=blue,
citecolor=blue,
urlcolor=blue,
menucolor=black
}

\title{Estimating Thematic Similarity of  Scholarly Papers \\with Their Resistance Distance in an Electric Network Model}
 \date{}

\author{Frank Havemann$^{1,\ast}$  \and Michael Heinz$^{1}$ \and Jochen Gl{\"a}ser$^{2}$ \and Alexander Struck$^{1}$}
\begin{document}
\maketitle
\begin{flushleft}

{1} 
Institut f{\"u}r Bibliotheks- und Informationswissenschaft, Humboldt-Universit{\"a}t zu Berlin, Berlin, Germany 
\\ 
 %\bf
{2} Zentrum Technik und Gesellschaft, Technische Universit{\"a}t Berlin, Berlin, Germany
\\ 

$\ast$ E-mail: Frank (dot) Havemann (at) ibi.hu-berlin.de
\end{flushleft}

\section*{Abstract}
We calculate resistance distances between papers in a nearly bipartite citation network of 492 papers and the sources cited by them. We validate that this is a realistic measure of thematic distance if each citation link has an electric resistance equal to the geometric mean of the number of the paper's references and the citation number of the cited source. % 

\section{Introduction}
It is often useful to be able to determine the thematic   similarity of two scholarly papers which is equivalent  to knowing their thematic distance in a     hypothetical space of concepts or in the genealogical tree   of knowledge. 
Modelling a scholarly paper by a set of terms and cited sources we face the problem that we cannot calculate the thematic similarity of two papers which themselves do not share terms respectively sources in their reference lists.
In bibliometric terms two such papers are neither bibliographically nor lexically coupled but it would not be adequate to assume that they  are totally unrelated because both types of lists, that of references and that of terms, are incomplete in general:
\begin{itemize}
 \item papers do not cite all of their intellectual ancestors,
\item very general descriptors are often not included in term lists.
\end{itemize}
In this paper we only discuss the case of citation networks of papers. Augmenting citation data with terms  improves  similarity estimation but we leave this opportunity for future work.

One solution of the problem of incomplete reference lists could be to search for all intellectual ancestors i.e.\ for indirect citation links between papers in the past. This would be a tedious task based on incomplete data because not all references of references are indexed in citation databases. Here we do not rely on indirect citation links in the past but on indirect connections in networks of papers and their cited sources in a time slice of one year. 

Earlier we tested whether thematic distances between papers could be estimated by the length of the shortest path between them in a one-year citation network~\shortcite{havemann2007mdr}.
Shortest-path length was also used by \citeN{botafogo1992sah} and by \citeN{egghe2003bcn} to measure compactness of unweighted networks. \citeN{egghe_measure_2003} generalised it to weighted graphs and applied it to small paper networks. A drawback of shortest-path length is its high sensitivity to the existence or nonexistence of single links which can act as shortcuts \cite{Mitesser2008mdr}.

Here we propose and test another solution, which takes all or at least the most important possible paths between two nodes into account: we calculate resistance distances between nodes \cite{klein_resistance_1993,tetali_random_1991}. To the best of our knowledge, resistance distance was not yet used for estimating the thematic similarity of papers.

We avoid time-consuming exact resistance computation by applying a fast approximate iteration method applied by \citeN{wu_finding_2004}. We also discuss other iterative approaches to the estimation of node similarity based on more than one path (s.\ sections \ref{method} and \ref{discussion}). 

In section \ref{validation} we validate that resistance is a realistic measure of thematic distance if each citation link has an electric conductance equal to the inverse geometric mean of the number of the paper's references and the citation number of the cited source.

\section{Method}
\label{method}
We use the nearly bipartite citation network of papers and their cited sources because projecting it on a one-mode graph of bibliographically coupled papers would reduce the information content of the data. The network is not fully bipartite because some papers are already cited in the year of their publication.

In the electric model we assume that each link has a conductance equal to its weight. We calculate the effective resistance between two nodes as if they  would operate as poles of an electric power source. Effective resistance has been proven to be a distance measure fulfilling the triangular inequation~\cite{klein_resistance_1993,tetali_random_1991}. 

One problem we have to solve before we can calculate distances is the delineation of the research field. It is not feasible and not necessary that the electric current between two papers flows through the total citation network of all papers published in the year considered. Field delineation should be done by an appropriate method for finding thematically coherent communities of papers \cite{fortunato2010community,havemann_identifying_2012}.

A second problem is the weighting of the network. In bibliometrics, the strength of citation links is often downgraded by dividing it by the number of references of the citing paper. We use a weighting of each link with the inverse geometric mean of its two nodes' degrees~\cite{havemann_evaluating_2012}: 
\begin{equation}
\label{link.weight} 
w_{ij} = \frac{A_{ij}}{\sqrt{k_i k_j}}.                                         \end{equation}
Then, for citation links, we take not only the number of references into account but also the number of citations the cited source recieves from the papers in the network. A citation link from a paper with many references to a highly cited source is weaker than a link from a paper with a short reference list to a seldom cited source.\footnote{ Such a weighting follows the same reasoning as the TF-IDF scheme in information retrieval.} 

A third problem is that an exact calculation of all resistance distances between $n$ nodes (e.g.\ papers and cited sources in a field) requires an inversion of an $n \times n$-matrix, a task of high complexity. Fortunately, we are only interested in similarities between papers and not between their many cited sources.  Furthermore, we need only approximations of similarity rather than exact values. Therefore we can apply a fast approximate iteration method applied by \citeN{wu_finding_2004} for community finding. We describe its details in Appendix \ref{appendix.1}.

This method is based on the fact that we know the effective resistance between two pole nodes in a network if we know the currents flowing from one of the two pole nodes to its neighbouring nodes. We can calculate these currents if we know the voltages of a pole's neighbours. From Kirchhoff's laws we know that---with the exception of the poles $p$ and $g$---a node's voltage is the average of its neighbours' voltages, more precisely the weighted average with link conductances as weights. 

If we start with all voltages equal to zero (except the positive pole's voltage $V_p=1$) we obtain the true voltages of all nodes by iteratively averaging voltages according to the formula  $V \gets F(p, g)V$, where $V$ is the voltage vector and $F(p, g)$ is the row normalised weighted adjacency matrix of the network but with the pole nodes'  row vectors filled with zeros with the exception of $F_{pp}=1$ (for details cf.\ Appendix \ref{appendix.1}).

There are other iterative approaches to the estimation of node similarity based on more than one path~\cite{leicht_vertex_2006,jeh_simrank:_2002}. Their  convergence can only be assured by introducing a parameter $< 1$ for downgrading the contributions of longer pathes. %or by breaking up the iteration after a fixed number of iterations .
Introducing an auxiliary parameter should be avoided unless its value could be estimated from theoretical consideration or from empirical data (cf.\ section \ref{discussion}).

\section{Experiment}
We experimented with community-finding algorithms on a connected citation network of 492 information science papers published in 2008 \shortcite{havemann2011identification}.\footnote{Source of raw data: Web of Science.}
In this sample we have identified three topics by inspection of titles, keywords, and abstracts \shortcite{havemann_identifying_2012}. Therefore, we also use it here to validate the measure of thematic distance of scholarly papers. 

\begin{figure}[t!] 
\begin{center} 
\includegraphics[width=3in]{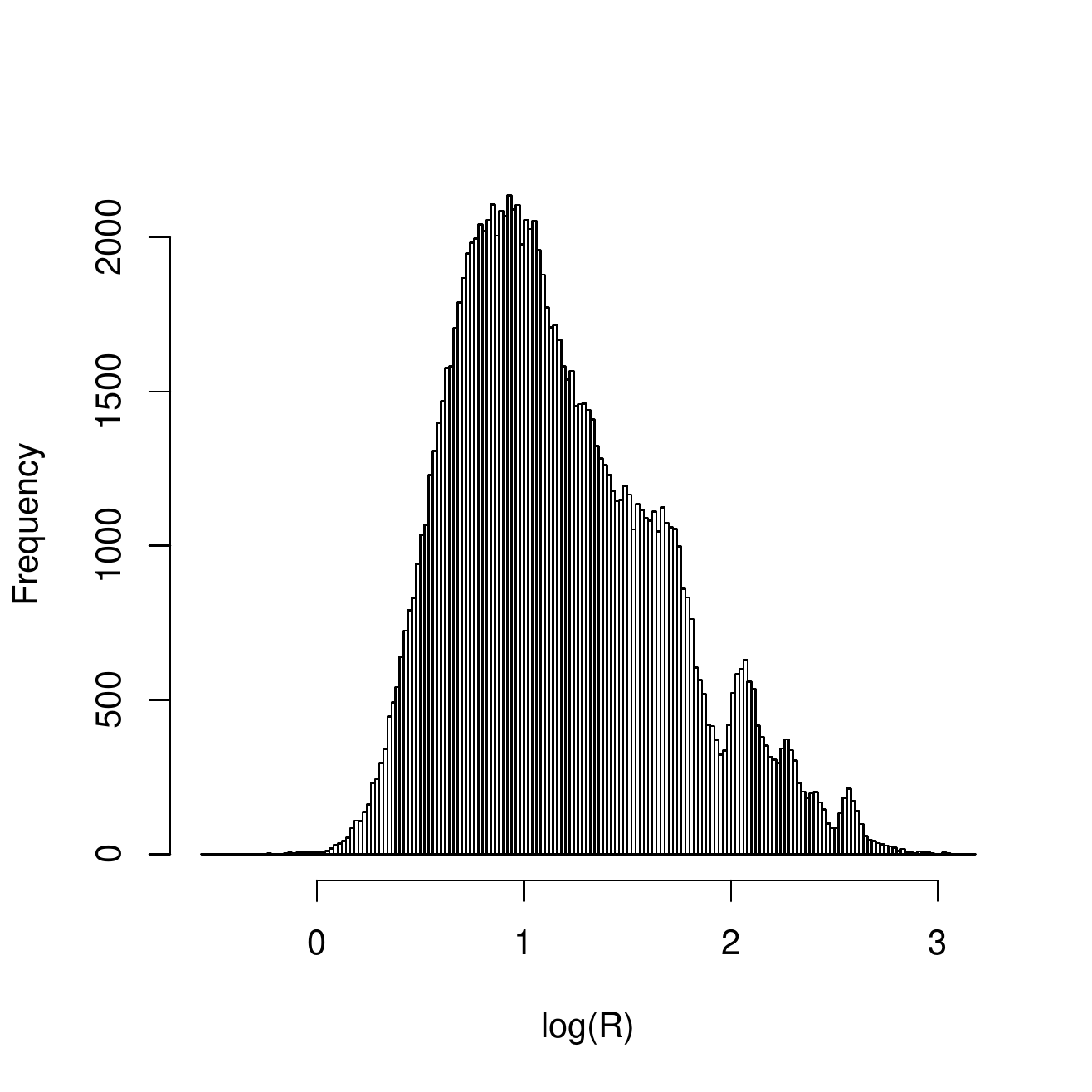} 
\end{center} 
\caption{Histogram of the logarithms of resistance distances between all 120,786 pairs of 492 papers. The distribution of $R$ is skewed, but that of $\log(R)$ rather symmetric.} 
\label{R-distribution} 
\end{figure}

The 492 papers cite 13,755 different sources and 21 other papers in the sample.
We analyse the nearly bipartite graph of papers and sources connected by 17,196 citation links. For the electric model we have to consider the graph to be undirected. We can drop all the 12,013 sources cited only once  because no current can flow through their citation links.  We cannot neglect the 15 papers cited only once. We weight the links according to equation \ref{link.weight} where $k_i$ is the degree of node $i$ after dropping the sources with only one citation.

The open-source C$^{++}$-program (written by Andreas Prescher) took about one hour to calculate the $492\cdot491/2 = 120,786$ distances with a maximal error of 0.1 (s.\ Figure~\ref{R-distribution}). If one only needs the distribution of distances it can be approximated by calculating distances of a random sample. Less then one third of distances (36,590) are needed to obtain an estimated standard error of the estimated population mean smaller than 0.01 (s.\ Appendix~\ref{appendix.2}).

\section{Validation}
\label{validation} 
 \begin{figure}[t!] 
\begin{center} 
\includegraphics[width=3in]{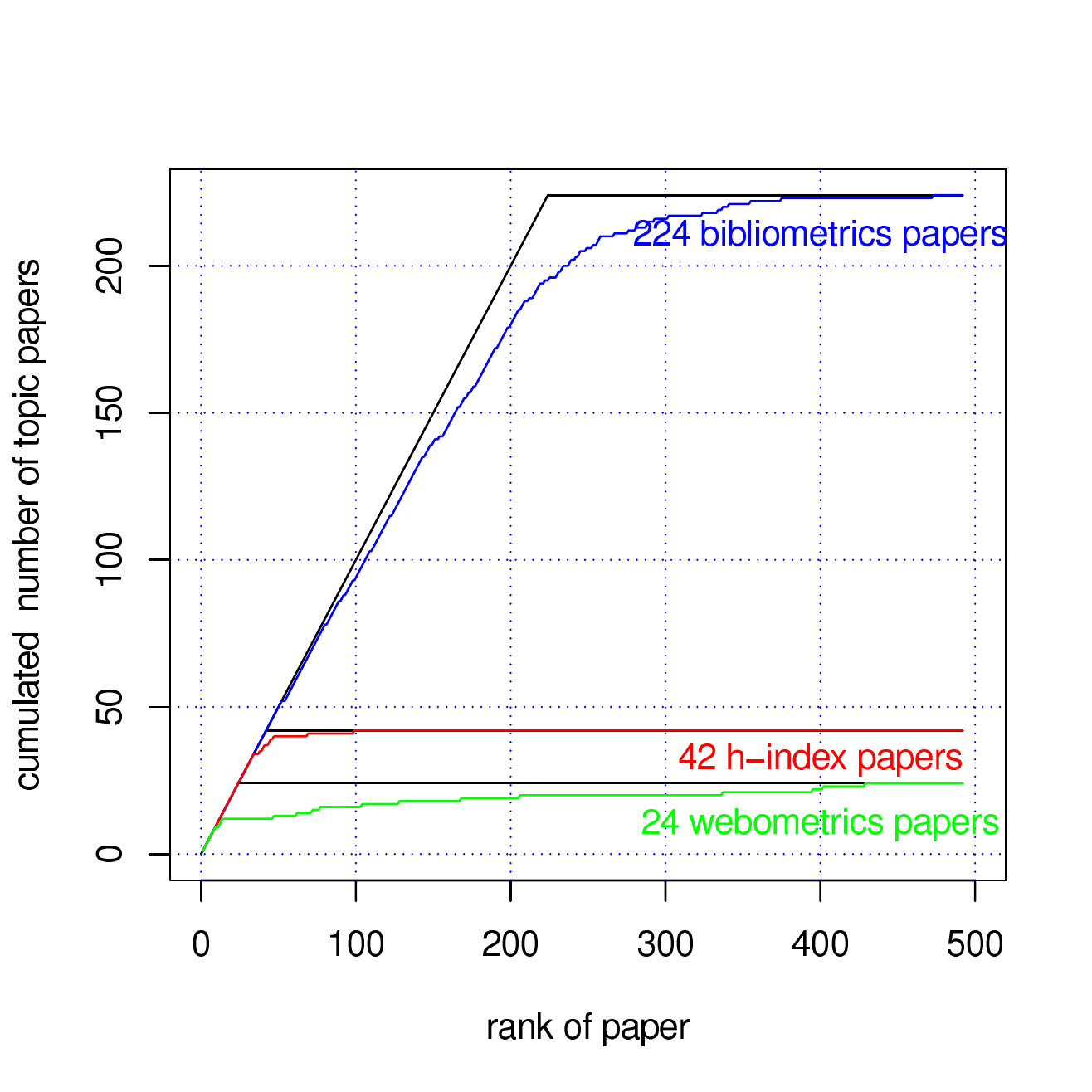} 
\end{center} 
\caption{Cumulated number of topic papers obtained from ranking all 492 papers according to their normalised median distance to papers of the three topics. The black lines represent the ideal cases, where all papers of a topic rank above other papers.} 
\label{R-recall} 
\end{figure}

In earlier research we had identified three overlapping topics in our network, named \textit{bibliometrics} (224 papers), \textit{Hirsch-index} (42 papers), and \textit{webometrics} (24 papers). %\shortcite{havemann_identifying_2012}
We validate the measure of thematic distance by ranking all papers according to the median distance to papers of a topic and expect the papers dealing with this topic at top ranks.
Because we have not classified really all papers dealing with the topics considered, the ranking with regard to thematic similarity cannot be perfect.

Another validation issue is that  on average resistance distances between high-degree nodes are smaller than between low-degree nodes because
%---in contrast to shortest-path lengths---resistance calculation is based on many possible paths between two nodes but 
all currents must flow through the immediate neighbours of the two nodes. The number of neighbours of a paper is the number of its references. More referenced sources suggest that the paper deals with more topics---at least in the discussion section. Thus, it is not an artifact of the measure that papers with many references have smaller distances to many other papers than papers with just a few references. In other words, they are often the central nodes in the graph.

Therefore we have to assure that the central nodes do not distort the ranking of nodes with regard to distances to a topic when we validate the measure. We correct for centrality by dividing the median distance of a paper to all topic papers by its median distance to all papers in the sample. The curves in Figure \ref{R-recall} show for the three predefined topics that indeed the topic papers have top ranks if we rank according to this ratio of medians.

This result is confirmed by a further test. We have used the resistance distances as an input for hierarchical clustering of papers. Ward clustering reconstructs the three topics with values of precision and recall  similar to the values we obtained with hierarchical link clustering \shortcite{havemann_identifying_2012}.

\section{Discussion}
\label{discussion}
There is  another approach to node similarity which also takes all possible paths between the nodes into account and also leads to an iterative matrix multiplication~\shortcite[and references of this paper]{leicht_vertex_2006}.\footnote{See  the paper by~\citeN{zhou_predicting_2009} for a discussion of further measures of node similarity.} 
It is based on a self-referential definition of node similarity inspired by self-referential influence definitions~\cite{Pinski1976citation,Brin1998anatomy}.

One advantage of the self-referential approach compared to the iterative resistance calculation is that one needs only one global iteration procedure to obtain all node similarities in one run. The most severe disadvantage we see is that the self-referential iteration does not converge unless an auxiliary multiplicative parameter $< 1$ is introduced which diminishes the weight which longer paths is given in the similarity measure. 

\shortciteN{leicht_vertex_2006} derive their iteration procedure by relating the number of observed paths of some length to the (approximated) number of expected paths between the two nodes. Such a relation to expectation is also necessary for the resistance approach if differences between distances have to be evaluated. A simple method is the one we apply for validation of our measure. We relate the observed to the median values of resistance distances.

If we want to obtain a similarity or distance measure which is comparable between different networks we have to relate resistance distances between nodes of a network to distances obtained in a null model of the network. The null model depends on the hypothesis we want to test with the measure of node similarity.

Applying their approach to the case of any two nodes $i$ and $j$ with distance 2, \shortciteN{leicht_vertex_2006} derive a similarity measure defined as the ratio of the number of common neighbours to the product $k_i k_j$ of their degrees (in contrast to the cosine similarity where this number is related to the square root of this product). If we estimate the current between two nodes which have common neighbours with the total current to the grounded pole $g$ from its neighbours after one iteration we get 
$$w_{pg}+\sum_{i = 1}^n \frac{w_{gi}w_{pi}}{w_i}$$ 
with $w_i = \sum w_{il}$ (cf.\ Appendix \ref{appendix.3}). For the network of a volume of papers and their cited sources there are only a few papers linked by a direct citation i.e.\ the first term nearly always vanishes: $w_{pg} = 0$. If the network is unweighted the similarity (measured with inverse distance) of two papers is then estimated by 
the sum of the inverse citation numbers $k_i$ of the sources cited by both papers
$$\sum_{i = 1}^n \frac{A_{gi}A_{pi}}{k_i},$$
a reasonable new absolute measure of bibliographic coupling where highly cited sources contribute less to the coupling strength than sources cited only by a few papers. With the weighting defined in equation \ref{link.weight} we obtain another measure of bibliographic coupling~(cf.\ Appendix \ref{appendix.3}):
$$\frac{1}{\sqrt{k_g k_p}} \sum_{i=1}^n \frac{ A_{gi} A_{pi} } { \sqrt{k_i} \sum_j A_{ij}/\sqrt{k_j} }.$$
Its denominator is equal to that of the cosine similarity and the common sources in the sum are weighted with the inverse product of the square root of their citation numbers and the sum over their citing papers weigthted with the inverse square root of their numbers of references. 

We do not propose to use this expression as a new similarity measure but argue that it is a reasonable relative measure of bibliographic coupling which downgrades the coupling strength of highly cited sources and downgrades the contribution to their citation numbers coming from papers citing many other sources. This confirms the weigthing we use here.

\section{Summary}
We have validated that resistance distance calculated in a citation graph is a realistic measure of thematic distance if each citation link has an electric resistance equal to the geometric mean of the number of the paper's references and the citation number of the cited source. % 

\section*{Acknowledgements}
%We thank Renaud Lambiotte and Steve Gregory for commenting on a draft of this paper.
This work is part of a project in which we develop methods for measuring the diversity of research. The project is funded by the German Ministry for Education and Research (BMBF). 
We thank Andreas Prescher for developing the fast C$^{++}$-program for the algorithm. %needed.
%We would like to thank all developers of \textbf{R}.\footnote{\url{http://www.r-project.org}}

%\section*{Author Contributions}
%Conceived the experiments: all authors.  Designed and performed the experiments: FH.  Analysed the data: FH. Wrote the paper: FH, JG. Discussed the text: all authors. 

\appendix
\section{Appendix} 
\subsection{Resistance Distance}
\label{appendix.1}

To calculate the total resistance between two nodes we apply the fast approximative method described by \citeN{wu_finding_2004}. 

To obtain the total resistance between any two nodes $p$ and $g$ we set the voltage $V_p$ of the positive pole $p$  to 1  and the voltage $V_g$ of the grounded pole $g$ to zero. Thus we get the
total tension $U = V_p - V_g = 1$. If we know the total current $I$ between the two poles 
 then we obtain the total resistance with $R = U/I = 1/I$.

If we know the voltages $V_i$ of the positive pole's adjacents $i$
 we obtain the total current $I$ by summing the currents
	$$I_{pi} = U_{pi}/R_{pi} = (V_p - V_i)/R_{pi} = (1 - V_i)/R_{pi} $$ 
 for all adjacents $i$.

Conductance $1/R_{ij}$ equals the link's weight $w_{ij}$.
 We therefore get for the total current $I$ between nodes $p$ and $g$ 
 	\begin{equation}
I = \sum_i I_{pi} = \sum_i w_{pi}(1 - V_i)  = w_p - \sum_i w_{pi}V_i  	                                \label{I}                                     \end{equation} 
 where $w_p = \sum_i w_{pi}$ is the weight of node $p$. We can also calculate the total current from the currents flowing into the grounded pole:
\begin{equation}
I = \sum_i I_{gi} = \sum_i w_{gi}V_i  	                                \label{I.g}.                                     \end{equation}

Each current $I_{ij}$ through link $(i, j)$ equals the voltage difference $U_{ij}$ of nodes $i$ and $j$ divided by the link's resistance $R_{ij}$:
 	$$I_{ij} = U_{ij}/R_{ij} = (V_i - V_j)/R_{ij}.$$
From Kirchhoff's laws we know that the sum of currents 
 flowing out of a node $i$ (which is not a voltage source) to its adjacents $j$ is zero:
 	$\sum_j I_{ij} =  0$,
 that means
	$$\sum_j(V_i - V_j)/R_{ij} = V_i w_i - \sum_j V_j/R_{ij} = 0.$$
 %where $w_i$ is $i$'s weight:
 %	$w_i = \sum_j w_{ij} = \sum_j 1/R_{ij}$.
 From this we obtain that the voltage of node $i$ is the weighted average of its adjacents' voltages:
 	\begin{equation}
V_i = \frac{1}{w_i}\sum_j  w_{ij}V_j. 	                              
\label{V}
\end{equation} 	
We obtain all the nodes' voltages by an iteration. For this, we turn equation \ref{V} into a command
 	$$V_i \gets \frac{1}{w_i} \sum_j  w_{ij} V_j, $$
 that means, in each iteration step, 
 we get the new voltage of a node by averaging the old voltages of the node's adjacents
 and expect that the algorithm converges.

If we introduce the weight matrix $W$ with row sums normalised to one by
$$W_{ij}=\frac{1}{w_i}\sum_j  w_{ij}$$
we can write the iteration command  as $V \gets WV$. 
Because the poles' voltages remain unchanged we use a matrix $F(p, g)$ instead of $W$. 
$F(p, g)$ is the row normalised weighted adjacency matrix of the network but with the pole nodes'  row vectors filled with zeros with the exception of $F_{pp}(p, g)=1$. 
%which we obtain be deleting the poles' rows in $W$. 

We only need the voltages of the positives pole's adjacents to obtain the total resistance beetween nodes $p$ and $g$ as $1/I$ with equation \ref{I}. During the iteration, we estimate these voltages. We consider the series of estimated voltages and observe that they cannot decrease. %The voltage ``flows'' from the positive pole into the network. 
This means, that the current $I$ estimated with equation \ref{I} does never increase and the total resistance $R=1/I$ does never decrease. From equation \ref{I} we obtain a lower bound of the true total resistance. Analogously, from equation \ref{I.g} we get an upper bound. Both bounds converge. We stop the iteration if the difference between both bounds becomes smaller than a small positive number $\epsilon$ which acts as a measure of precision needed for the analysis.

\subsection{First Approximation for Poles with Common Neighbours}
\label{appendix.3}
We start with voltages $V_i=0, \forall i \ne p$ and $V_p=1$. The first iteration results in voltages
\begin{equation}
 V_i(1) = \dfrac{1}{w_i}\sum_j w_{ij}V_j(0)=\dfrac{w_{ip}}{w_i}, \forall i \ne p,g.
\end{equation} 
The current reaching the grounded pole is then
\begin{equation}
 I_g(1) = \sum_i w_{gi}V_i(1) = w_{pg} + \sum_i \dfrac{w_{gi}w_{ip}}{w_i}.
\end{equation} 
If positive pole $p$ and grounded pole $g$ have a graph distance of two hops then  $w_{pg}=0$.

In the unweighted case $w_{ij}=A_{ij}$ and 
$$ I_g(1) = \sum_i \dfrac{A_{gi}A_{ip}}{k_i}.$$
If we weight according to equation \ref{link.weight} we obtain
\begin{equation}
 I_g(1) = \sum_i \dfrac{A_{gi}A_{ip}}{ \sqrt{k_gk_i}\cdot\frac{1}{\sqrt{k_i}}\sum_j\frac{A_{ij}}{\sqrt{k_j}}\cdot  \sqrt{k_ik_p}}
\end{equation} 
or
$$I_g(1) =\frac{1}{\sqrt{k_g k_p}} \sum_{i=1}^n \frac{ A_{gi} A_{pi} } { \sqrt{k_i} \sum_j A_{ij}/\sqrt{k_j} }.$$

\subsection{Distances of a Random Sample}
\label{appendix.2} 
If we do not need distances between all $|P|$ papers but only the form of their distribution we can avoid to calculate all $|P|(|P|-1)/2$ distances.  In this case, we order all paper pairs randomly. Then the first $n$ distances are a random sample from all  distances. The standard error $S_R$ of the average resistance $R$ is then given by the square root of
\begin{equation*}
S_R^2 = \frac{N-n}{(N-1)(n-1)n} \left[ \sum_{i=1}^n R_i^2 - \frac{1}{n}\left(\sum_{i=1}^n R_i\right)^2\right].
\end{equation*} 
We stop calculating resistance distances if standard error $S_R$ is smaller than $\epsilon/10$ for the last ten random samples. We can choose a relative large
$\epsilon$ for precision of each single resistance because the average remains precise even if the averaged values are rounded. Both sums in the formula can be updated easily by adding the new terms to the last values of the sums.

The formula  for $S_R^2$ can be derived from
\begin{equation}
 S_R^2 = \frac{N-n}{(N-1)n}S^2,
\end{equation} 
where the variance of distances $S^2$ can be estimated by the variance of the sample $s^2$ according to
\begin{equation}
S^2 = \frac{n}{n-1}s^2. 
\end{equation} 
We have 
\begin{equation*}
s^2 =  \frac{1}{n} \sum_{i=1}^n(R_i - R)^2=\frac{1}{n} \left[ \sum_{i=1}^nR_i^2 -\frac{1}{n}\left(  \sum_{i=1}^n R_i\right) ^2\right],
\end{equation*} 
leading to the formula for $S_R^2$.
\bibliography{informetrics}

\end{document}